\PassOptionsToPackage{protrusion=false, expansion=false}{microtype}
\documentclass[manuscript]{acmart}

\usepackage{microtype}
\usepackage{subfigure}
\usepackage{booktabs} % for professional tables
\usepackage{bbding}
\usepackage[most]{tcolorbox}  % Place in preamble

\usepackage{hyperref}

\usepackage{pslatex}
\usepackage{soul}
\usepackage{csvsimple}

\usepackage[textsize=tiny]{todonotes}

\usepackage{listings}
\usepackage{xcolor}
\definecolor{codegreen}{rgb}{0,0.6,0}
\definecolor{codegray}{rgb}{0.5,0.5,0.5}
\definecolor{codepurple}{rgb}{0.58,0,0.82}
\definecolor{backcolour}{rgb}{0.95,0.95,0.92}

\lstdefinestyle{cstyle}{
    language=C,
    basicstyle=\ttfamily\small,
    backgroundcolor=\color{backcolour},   
    commentstyle=\color{codegreen},
    keywordstyle=\color{magenta},
    numberstyle=\tiny\color{codegray},
    stringstyle=\color{codepurple},
    breaklines=true,
    captionpos=b,                    
    keepspaces=true,                 
    showspaces=false,                
    showstringspaces=false,
    showtabs=false,                  
    tabsize=2,
    moredelim=[is][\color{red}]{@@}{@@},
    }

\lstdefinestyle{csvstyle}{
    basicstyle=\ttfamily\footnotesize,
    columns=fullflexible,
    breaklines=true,
    postbreak=\mbox{\textcolor{red}{$\hookrightarrow$}\space},
    frame=single,
    captionpos=b,
    showstringspaces=false,
    moredelim=[is][\textbf]{**}{**}
}

\lstdefinestyle{cotstyle}{
    basicstyle=\small\sffamily,
    language={},
    moredelim=[is][\color{red}]{@@}{@@}%,
}

\usepackage{mdframed}
\newmdenv[
    linecolor=black,
    linewidth=2pt,
    backgroundcolor=gray!20,
    frametitlebackgroundcolor=gray!20,
    frametitlerule=true,
    roundcorner=10pt
]{mybox}

\definecolor{babyblue}{rgb}{0.54, 0.81, 0.94}

\title[Seeking Specifications]%
{Seeking Specifications: The Case for Neuro-Symbolic Specification Synthesis}
\thanks{This work was supported by the Wallenberg AI, Autonomous Systems and Software Program (WASP), funded by the Knut and Alice Wallenberg Foundation.}

\author{George Granberry}
\orcid{0009-0005-4628-5464}
\affiliation{%
  \institution{Chalmers University of Technology and University of Gothenburg}
  \department{Computer Science and Engineering}
  \city{Gothenburg}
  \country{Sweden}
}
\email{georgegr@chalmers.se}

\author{Wolfgang Ahrendt}
\orcid{0000-0002-5671-2555}
\affiliation{%
  \institution{Chalmers University of Technology and University of Gothenburg}
  \department{Computer Science and Engineering}
  \city{Gothenburg}
  \country{Sweden}
}
\email{ahrendt@chalmers.se}

\author{Moa Johansson}
\orcid{0000-0002-1097-8278}
\affiliation{%
  \institution{Chalmers University of Technology and University of Gothenburg}
  \department{Computer Science and Engineering}
  \city{Gothenburg}
  \country{Sweden}
}
\email{moa.johansson@chalmers.se}

\begin{document}
%


%

%\icmlkeywords{specification, synthesis, LLM, formal methods}

\begin{abstract}
This work is concerned with the generation of formal specifications from code, using Large Language Models (LLMs) in combination with symbolic methods. Concretely, in our study, the programming language is C, the specification language is ACSL, and the LLM is Deepseek-R1. In this context, we address two research directions, namely the specification of intent vs. implementation on the one hand, and the combination of symbolic analyses with LLMs on the other hand.
For the first, we investigate how the absence or presence of bugs in the code impacts the generated specifications, as well as whether and how a user can direct the LLM to specify intent or implementation, respectively.
For the second, we investigate the impact of results from symbolic analyses on the specifications generated by the LLM. The LLM prompts are augmented with outputs from two formal methods tools in the Frama-C ecosystem, Pathcrawler and EVA. We demonstrate how the addition of symbolic analysis to the workflow impacts the quality of annotations.
\end{abstract}

\maketitle

\section{Introduction}
The field of specification synthesis offers a possible solution to the inherent complexities involved in creating and maintaining specifications for software verification. Creating useful specifications demands a deep understanding of both the specification language and the verification process, which can often be as intricate, if not more so, than the software to be verified. This complexity poses a significant barrier \cite{davis2013study,tyler2021formal}, especially in dynamic environments where frequent updates and refactoring are the norm. Maintaining an accurate alignment between ever-evolving code and its specifications can become a cumbersome and error-prone process.

Specification synthesis potentially alleviates these concerns by automating the generation and adaptation of specifications. Instead of requiring developers to manually write detailed specifications -- a task that can be both time-consuming and susceptible to human error -- specification synthesis aims to infer and edit specifications directly from the codebase and associated context. The goal is to transform specifications into convenient guardrails that provide valuable insights and guidance to programmers, rather than chores performed at the end of the software pipeline.

Approaches to generating specifications typically employ a range of symbolic techniques, encompassing static as well as dynamic analyses \cite{lathouwers2024survey}. For instance, Daikon \cite{ernst2007daikon}, a widely recognised tool in dynamic analysis, infers properties by observing program behaviour at runtime. On the other hand, static analysers deduce properties based on the program's structure without executing it. Despite their precision, the primary limitation of these methods is their rigidity. Symbolic techniques are constrained by a limited range of expressible properties and typically specialise in specific types of analyses which restricts their flexibility in adapting to diverse verification needs.

On the other side of specification synthesis techniques are machine-learning-based Natural Language Processing (NLP)~\cite{blasi2018translating} and Large Language Models (LLMs) \cite{brown2020language}. NLP tools specialise in understanding human language, such as comments, while LLMs stand out for their flexibility and creativity in when dealing with arbitrary inputs. These models can theoretically generate any specification that can be articulated in their associated language, provided that they are appropriately trained and given the right prompts.

However, this strength also introduces a significant challenge: the large range of potential specifications LLMs can produce often includes outputs that may not be practically useful or even plainly wrong. While an LLM can generate a wide array of specifications, the lack of inherent direction means that there is no guarantee that the generated specifications will be relevant or valuable for specific verification tasks. This challenge has led users of LLM-based synthesis to utilize \emph{prompt engineering} \cite{white2023prompt} techniques in order to increase the likelihood of the LLM to produce outputs that align with their objectives. Despite the seeming effectiveness of this practice across general LLM usage, little research has been done to figure how these prompting techniques are utilized by LLMs when it comes to specification synthesis.

A fundamental conceptual issue in the context of specification synthesis is that specifications synthesized directly from code may reflect the implemented behavior rather than the intended behavior. Depending on the purpose of the specification, this can be seen as problematic in contexts where correctness cannot simply be assumed, and synthesized specifications risk formalizing unintended behavior, potentially obscuring bugs instead of revealing them.

In this paper, we demonstrate that specification synthesis is a viable and effective practice using current-day tools. First, we explore how LLMs handle the conceptual challenge of distinguishing between implemented and intended behavior when synthesizing specifications. Afterwards, we propose a complementary integration of existing symbolic analysis tools with LLM-based specification synthesis, combining the precision of symbolic methods with the flexibility and expressive power of LLMs.

In our experiments, we observed that LLMs - particularly those with advanced reasoning capabilities - were resilient against the implementation-versus-intent problem. These models were almost always able to identify bugs and target \textbf{intended} behavior rather than the implemented behavior when provided incorrect or buggy code. Furthermore, their reasoning capabilities enabled effective bug detection \textbf{even with no context provided}. We also found that prompts can be adjusted to explicitly acknowledge the possibility of bugs, thereby further guiding the LLM toward producing specifications that reflect the developer's intent rather than the implemented behavior.

In addition, we find that incorporating results from symbolic analysis into prompts significantly impacts both the quantity and content of generated specifications. To evaluate this effect, we focus on tooling from the Frama-C ecosystem, which provides formal analysis tools for C programs. We express specifications using the ACSL specification language and leverage two tools: PathCrawler, a structural test case generator that produces input/output examples, and EVA, a static analyzer that performs value analysis. In general, augmenting the prompts with output symbolic tools led to fewer annotations overall, but increased the relevance and focus of the resulting specifications. Symbolic input influenced the type of annotations as well: PathCrawler encouraged the generation of more structured and abstract postconditions, while EVA led to an increase in preconditions, directly reflecting its emphasis on identifying runtime errors. Together, these results highlight the adaptability of LLMs to integrate structured, symbolic information into their specification synthesis process.

\begin{tcolorbox}[colback=gray!10, colframe=black, title=Research Questions]
This paper investigates the following research questions:

\begin{enumerate}
    \item How does LLM-based specification synthesis from code handle buggy code?
    \item How can existing symbolic methods be used to influence and improve LLM-based specification synthesis methods?
\end{enumerate}
\end{tcolorbox}

This article extends and modifies \cite{granberry2024specify} in several ways. First, we replace OpenAI's GPT-4 with Deepseek-R1 and re-run all experiments using this newer model. Second, we significantly expand the section on implementation versus intent by introducing a larger and more diverse dataset designed to test that distinction. Finally, we incorporate Deepseek-R1's more detailed reasoning output into our qualitative analysis, allowing for a more comprehensive understanding of how the model arrives at its specifications.

The structure of the paper is as follows. First, we introduce the experimental setup, covering the tools, datasets, and prompt strategies used. Second, we present an investigation into the implementation-versus-intent distinction, including a dataset constructed to test how LLMs handle buggy or misleading implementations. Finally, we investigate neuro-symbolic prompting experiments, investigating the effect of adding the output of symbolic analyses to the prompts when generating specifications from code with Deepseek-R1.
%to assess how symbolic information affects specification synthesis in the context of this newer model.

\section{Tools and Languages}
%\section{Methods and Tools}
%Describe what LLM we use, and explain briefly what Farma-C is, what annotations look like and specifically introduce Pathcrawler and EVA.

\subsection{Frama-C and ACSL}
The Frama-C ecosystem is an open-source suite of tools designed for the analysis of the source code of software written in C \cite{Farma-C,frama-cBook2024}. It integrates various static and dynamic analysis techniques to evaluate the correctness, safety, and security of C programs. It also supports the specification language ACSL \cite{baudin2008acsl,e_acsl_reference}, which is used to formulate \emph{contracts} consisting of, among others, preconditions – assumptions on the input and prestate of a function – and postconditions – requirements on the output and poststate of a function.
These contracts, examples of which can be seen both in Sections~\ref{experiment1} and~\ref{experiment2} , provide a clear and formal framework for understanding and
verifying a function’s behaviour. Other ACSL annotations commonly used are \emph{assertions} - stating a condition that needs to be true at some point in execution - and \emph{loop invariants} which specify conditions that need to be maintained by every iteration of a loop.

\subsection{Automated Test Generation: Pathcrawler}
The PathCrawler tool is designed for the automated testing of C programs \cite{williams2005pathcrawler}. Its primary function is to generate and execute test inputs for C code, with a particular focus on achieving high code coverage. Employing a technique known as concolic testing \cite{concolicTestingC05} -- using a combination of concrete and symbolic execution -- Pathcrawler efficiently explores different execution paths in the program. First it generates test inputs and then executes them, providing valuable information from the execution results across a broad spectrum of program paths. In addition, PathCrawler allows users to incorporate a test oracles, classifying the outcome of every test case of some function. However, we want to highlight that we did not make such oracle implementations available to the LLM when asking it to generate specifications. Generic oracles can be seen as executable specifications, and would have diluted the significance of our experiments. Instead, we only include input/output pairs, with the non-generic verdict. An example of output from PathCrawler can be seen in Fig.~\ref{pc:csv}.
\begin{figure}[hbpt]
\begin{lstlisting}[style=csvstyle]
input_l,input_table[0],input_table[1],output,verdict
2,0,0,,success
2,0,73,73,0,success
\end{lstlisting}
\caption{Example PathCrawler input/outputs for BubbleSort}
\label{pc:csv}
\end{figure}

\subsection{Value Analysis: EVA}
The EVA static analyzer uses abstract interpretation to approximate a set of possible values that program variables can take to avoid certain runtime errors \cite{EVA}. By doing so, it can
identify a range of potential issues, such as division by zero, buffer overflows, null
pointer dereferences, and arithmetic overflows. EVA’s analysis helps in ensuring
that the code behaves correctly across all possible execution paths and input
values. EVA is designed to respect, and work with, ACSL annotations when they are present. For an example of an report ouput from EVA refer to Fig.~\ref{eva_report}.

% the symbol "≤" did not compile. Used #\textcolor{red}{$\leq$}# instead:

\begin{figure}[hbpt]
\begin{lstlisting}[style=cotstyle,escapechar=\#]
...
@@[eva:alarm] temp_files/tmphpdyn83w/eva_temp.c:8: Warning: 
  signed overflow. assert -2147483648 #\textcolor{red}{$\leq$}# x * 2;
[eva:alarm] temp_files/tmphpdyn83w/eva_temp.c:8: Warning: 
  signed overflow. assert x * 2 #\textcolor{red}{$\leq$}# 2147483647;
[eva:alarm] temp_files/tmphpdyn83w/eva_temp.c:9: Warning: 
  signed overflow. assert -2147483648 #\textcolor{red}{$\leq$}# v - y;
[eva:alarm] temp_files/tmphpdyn83w/eva_temp.c:9: Warning: 
  signed overflow. assert v - y #\textcolor{red}{$\leq$}# 2147483647;
[eva:alarm] temp_files/tmphpdyn83w/eva_temp.c:11: Warning: 
  out of bounds write. assert \valid(tab + 2);
[kernel] temp_files/tmphpdyn83w/eva_temp.c:11: Warning: 
  all target addresses were invalid. This path is assumed to be dead. @@
[eva] done for function testme
[eva] temp_files/tmphpdyn83w/eva_temp.c:11: 
  assertion 'Eva,mem_access' got final status invalid.
[eva] ====== VALUES COMPUTED ======
[eva:final-states] Values at end of function testme:
  NON TERMINATING FUNCTION
[eva:summary] ====== ANALYSIS SUMMARY ======
  ----------------------------------------------------------------------------
  1 function analyzed (out of 1): 100% coverage.
  In this function, 4 statements reached (out of 14): 28% coverage.
  ----------------------------------------------------------------------------
  Some errors and warnings have been raised during the analysis:
    by the Eva analyzer:      0 errors    0 warnings
    by the Frama-C kernel:    0 errors    1 warning
  ----------------------------------------------------------------------------
  5 alarms generated by the analysis:
       4 integer overflows
       1 invalid memory access
  1 of them is a sure alarm (invalid status).
\end{lstlisting}
\caption{Example EVA report}
\label{eva_report}
\end{figure}

\subsection{LLM and Prompts}
We have chosen to use Deepseek-R1\footnote{\url{https://platform.deepseek.com}} \label{deepseek} (deepseek-reasoner version 2025-01-20) as our LLM for generating specifications. We ran this experiment using GPT-4 in our previous work~\cite{granberry2024specify} but chose to switch to Deepseek-R1, not only for its competitive pricing, but also for its more extensive reasoning output.

We prompt Deepseek-R1 with a C program, providing instructions for how to generate ACSL annotations in a step-by-step manner. We also include a few examples of valid ACSL annotations in the prompts, leveraging a form of “few-shot learning”~\cite{brown2020language} to guide the model. Unlike previous versions of this paper, we did not choose to include "Chain-of-thought"~\cite{wang2023chain} prompting techniques in the prompt as reasoning models such a Deepseek-R1 perform this by default. Our "baseline" prompt, which is used to generate specifications without the aid of symbolic analysis, can be seen in Fig.~\ref{prompt:baseline} on page ~\pageref{prompt:baseline}.

\begin{figure}[hbpt]
\begin{lstlisting}[style=cotstyle]
You are a LLM that takes the following inputs and returns a C program annotated with ACSL annotations.

Inputs:
1. A C program with no ACSL annotations

GOALS:
1. Describe any abstract properties that could be represented as ACSL annotations
2. Generate ACSL annotations that describe the functional behavior of the program based on your analysis of the program
3. Returning a program with no annotation is not a valid solution
4. Do not edit the C code, only add annotations
5. Do not add any annotations inside of the function body
6. Do not generate loop invariants
7. Do not skip any code in the returned solution to make it shorter.

ANNOTATION EXAMPLES:

Examples 1 (single annotation):
/*@ requires low >= 0 && high <= 9; */

Example 2 (block annotation style):
//Only use this style for function headers. Do not use blocks for multiple annoations in the function body
/*@ 
  @ requires low >= 0 && high <= 9;
  @ requires elem >= 0 && elem <= 9;
*/

FORMAT INSTRUCTIONS:

First describe your reasoning behind the added annotations

Return the annotated c code wrapped in markdown
```c
...
```
 
START OF INPUT:

```c
{program_str}
```
\end{lstlisting}
\caption{Prompt used for generating ACSL annotations}
\label{prompt:baseline}
\end{figure}

Expanding upon the baseline prompt, in Sect.~\ref{experiment2} we augment it with outputs from the Pathcrawler and EVA tools to further guide the model. These two prompts can be found in Fig.~\ref{prompt:pc} and~\ref{prompt:eva} on pages~\pageref{prompt:pc} and~\pageref{prompt:eva} respectively.

\begin{figure}[hbpt]
\begin{lstlisting}[style=cotstyle]
You are a LLM that takes the following inputs and returns a C program annotated with ACSL annotations.

Inputs:
1. A C program with no ACSL annotations
2. A CSV file that represents input/output pairs that result from running the Frama-C Pathcrawler tool

GOALS:
1. Descrbe any abstract or functional properties that you can reason about from the pathcrawler output
2. Describe any abstract or functional properties based on the code
3. Generate ACSL annotations that describe the functional behavior of the program based on both your analysis of of the program as well as the pathcrawler output
4. Returning a program with no annotation is not a valid solution
5. Do not edit the C code, only add annotations
6. Do not add any annotations inside of the function body
7. Do not generate loop invariants
8. Do not skip any code in the returned solution to make it shorter.

ANNOTATION EXAMPLES:

Examples 1 (single annotation):
/*@ requires low >= 0 && high <= 9; */

Example 2 (block annotation style):
//Only use this style for function headers. Do not use blocks for multiple annoations in the function body
/*@ 
  @ requires low >= 0 && high <= 9;
  @ requires elem >= 0 && elem <= 9;
*/

FORMAT INSTRUCTIONS:
First describe your reasoning behind the added annotations
Return the annotated c code wrapped in markdown
```c
...
```
START OF INPUT:
```c
{program_str}
```
```csv
{pathcrawler_str}
```
\end{lstlisting}
\caption{Prompt used for generating ACSL annotations with PathCrawler as additional context}
\label{prompt:pc}
\end{figure}

\begin{figure}[hbpt]
\begin{lstlisting}[style=cotstyle]
You are a LLM that takes the following inputs and returns a C program annotated with ACSL annotations.

Inputs:
1. A C program with no ACSL annotations
2. A report generated by the Frama-C EVA static analysis tool

GOALS:
1. Descrbe any abstract or functional properties that you can reason about from the EVA report
2. Describe any abstract or functional properties based on the code
3. Generate ACSL annotations that describe the functional behavior of the program based on both your analysis of of the program as well as the EVA report
4. Returning a program with no annotation is not a valid solution
5. Do not edit the C code, only add annotations
6. Do not add any annotations inside of the function body
7. Do not generate loop invariants
8. Do not skip any code in the returned solution to make it shorter.

ANNOTATION EXAMPLES:

Examples 1 (single annotation):
/*@ requires low >= 0 && high <= 9; */

Example 2 (block annotation style):
//Only use this style for function headers. Do not use blocks for multiple annoations in the function body
/*@ 
  @ requires low >= 0 && high <= 9;
  @ requires elem >= 0 && elem <= 9;
*/

FORMAT INSTRUCTIONS:

First describe your reasoning behind the added annotations

Return the annotated c code wrapped in markdown
```c
...
```

START OF INPUT:
```c
{program_str}
```

```
{eva_str}
```

\end{lstlisting}
\caption{Prompt used for generating ACSL annotations with EVA as additional context}
\label{prompt:eva}
\end{figure}

\section{Implementation vs. Intent}
\label{experiment1}
A core criticism of specification synthesis is that it risks producing incorrect specifications when the source code contains bugs. One might argue that this issue is not critical since users should be able to identify faulty specifications just as they would identify faulty code and revise accordingly. However, modern specification languages such as ACSL are expressive programming languages in their own right. As such, users may be just as prone to overlooking subtle errors in specifications as they are to overlooking bugs in code. This creates a potential for overconfidence in specifications that merely formalize flawed behavior, rather than revealing it. For this reason, the distinction between implementation and intent becomes a central concern. If LLM-based synthesis tools cannot recover the intended intent of a program from buggy code, then the synthesized specification may reinforce errors rather than assist in their detection.

LLMs have opened up new pathways for addressing the challenges of specification synthesis.
%Unlike traditional symbolic tools, LLMs have the potential to reason abstractly and recognize patterns.
In this section, we perform a qualitative analysis of the reasoning processes generated by Deepseek-R1 to explore how these capabilities manifest in practice. By examining how the model responds to mismatches between implementation and intent, we aim to understand whether and how LLMs can identify discrepancies, infer programmer intent, and produce specifications that go beyond simply restating buggy behavior.

Specifically, we aim to answer the following questions:
\begin{enumerate}
    \item How capable are LLMs at identifying the intent of a program?
    \item How reliant are they on cues such as function names and comments?
    \item Is an LLM able to recognize subtle bugs in programs?
    \item If a bug is recognized, will the LLM generate a specification that mirrors the buggy implementation or one that follows the intent of the program?
\end{enumerate}

\subsection{Dataset}
The dataset used in this study is publicly available at \url{https://github.com/ggranberry/intent_dataset}. We collected programs for a test suite called \emph{intent\_tests}, composed of 50 C programs, designed to investigate how LLMs manage conflicts between implementation and intention. This suite is structured into four sections:

The first section, \emph{Basic}, contains 20 programs commonly taught in undergraduate computer science curricula, including binary search, duplicate identification, queue insertion, linked-list append, string reversal, and palindrome checking. These serve as a benchmark on standard, widely recognized algorithms.

The second section, named \emph{Famous}, includes 10 more complex but still well-known programs, such as Duff's Device and Kahan's Summation Algorithm. These programs present a more challenging scenario for an LLM in which it is familiar with the provided program but might have trouble generating an appropriate specification.

In the third section, \emph{Mirror}, we include 10 intentionally brief programs, mostly consisting of just one or two lines of code. These are specifically designed so their implementations can be straightforwardly and precisely mirrored into ACSL specifications. The purpose of this section is to test how tightly coupled the specifications are to the implementation.

\subsection{Generating Annotations}
\label{section:generating_annotations}
The variability of LLMs like Deepseek-R1 can be adjusted via its "temperature" setting which controls the level of determinism during generation. As we are interested in exploring what the \textit{average} specification generated by a given prompt is, we choose to generate three distinct specifications for each program (and prompt) within our test suite, repeating the steps above with a temperature setting of 0.7. This approach allows us to capture a spectrum of possible specifications, while not being too economically costly.

Given that we are using the \texttt{deepseek-reasoner} model, each generated response consists of two parts: the ACSL specification and an accompanying chain-of-thought reasoning provided by the model prior to generating the specification. Having direct access to the LLM’s reasoning process is particularly beneficial, as it provides insight into the internal logic employed by the model. Specifically, it allows us to pinpoint nuance such as determining whether the LLM recognizes familiar algorithms, discerns the intended functionality of the provided programs, or identifies the subtle bugs we deliberately introduced.

Loop invariants are key for functioning specifications. However, generating loop invariants is not the focus of this paper. The generation of loop invariants with LLMs calls for sophisticated prompts dedicated to this task \cite{kamath2023finding,flanagan2001annotation}. Rather, we focus on function contracts -- particularly preconditions, postconditions, and assigns clauses -- and our prompts are designed to generate these while leaving out loop invariants.
 
\subsection{Evaluation}
\label{sec:intent_eval}
Evaluating specifications is challenging due to the absence of a definitive specification for any given program. Different users often have varying priorities and perspectives on which properties are worth verifying, making the notion of a definitive specification subjective. Similarly, a specification might be logically correct, but more or less trivial with respect to the program at hand, in which case it provides little value.

In light of these challenges, our evaluation methodology does not attempt to benchmark the generated specifications against any predefined gold standard, nor does it aim to determine the optimal approach to creating specifications. Instead, our focus is on identifying the behaviours and patterns that emerge when generating specifications with controlled prompting and program edits.

While the generated specifications themselves are interesting, we found that the reasoning process that lead to the specification to be much more informative. To this point, we primarily focused on analysis of the \textbf{reasoning output} generated from the different variations of the \emph{Intent Tests} suite. In these tests, we treat the programs themselves as controlled variables, systematically introducing buggy variations to assess how capable the LLM at pinpointing and specifying user intent in code.

We generated specifications for four variations of the \emph{Intent Tests} suite:

\begin{enumerate}
    \item \textbf{Baseline Set}: The original, unmodified set of programs, which we consider correct to the best of our knowledge. This serves as our baseline for comparing the effects of our subsequent modifications.
    \item \textbf{Buggy Set}: We introduced subtle bugs into each of the programs. These errors were intentionally designed to be syntactically minimal at the same time as they significantly impact the semantic correctness of the programs.
    \item \textbf{Anonymized Set}: We anonymized the function names in the baseline set, replacing meaningful identifiers with generic placeholders to obscure the original intent from the LLM.
    \item \textbf{Buggy Anonymized Set}: This variation combined both anonymization of function names and the introduction of the same subtle bugs used in the Buggy Set.
\end{enumerate}

\subsection{Identifying Intent}
Deepseek-R1 was overwhelmingly successful at identifying the intent behind the programs provided. Across nearly 600 specification generations, the model correctly inferred the intended functionality for nearly every program, including those with injected bugs and anonymized function names.

The 3 exceptions in which the intent was not able to be identified occurred in programs that were both buggy and anonymized. In these few cases, the injected bugs altered the program semantics significantly enough that the program more resembled another coherent program.

\subsection{Reliance on Identifiers}
In the unmodified version of the \emph{Intent Tests}, programs were presented with descriptive and meaningful function names. This setup naturally provides strong cues that can make the task of identifying program intent relatively trivial for an LLM. However, in our experiments, anonymizing these function names had surprisingly little effect on the model’s performance. In fact, Deepseek-R1 seemed to be slightly better at identifying bugs with anonymized function names.

In the reasoning outputs, we consistently observed that the LLM was either able to immediately recognize the program or experienced an "Aha!" moment partway through the reasoning process. This indicates that the model's recognition capabilities are driven more by its understanding of the program itself rather than simple reliance on identifier names.

\begin{lstlisting}[style=cotstyle]
Wait, this looks familiar. Oh right! This is the fast inverse square root algorithm, but modified.
\end{lstlisting}
\label{listing:inverse_square_root}

\subsection{Identifying Bugs}
Deepseek-R1 was also effective at identifying subtle bugs in code, even when not explicitly prompted to search for them. As shown in Table~\ref{tab:bugs-noticed-corrected}, the model was able to detect bugs in the majority of the programs provided.

However, the model was noticeably less effective in detecting bugs in the \emph{Famous} category of the test suite. One possible interpretation for this is that the programs in this group are more complex than those in the other categories, making them more difficult to reason about. Another insight, revealed through examination of the reasoning traces, is that the model often struggled to identify bugs in these cases because it had difficulty constructing meaningful specifications.

For instance, in the case of Kahan's Summation Algorithm, which is designed to reduce precision loss in floating-point summation, the model immediately identified the purpose of the program. However, it then spent most of its reasoning phase attempting to determine which properties could be usefully specified for such a program. This effort left little time or attention for analyzing the structure or behavior of the code itself, thereby reducing its ability to detect the bug. This pattern was repeated in other complex examples, where the effort to determine an appropriate specification overshadowed reasoning about program correctness.

\begin{table}[H]
\centering
\begin{tabular}{l r}
\toprule
\textbf{Test Suite} & \textbf{Bugs Noticed} \\
\midrule
Basic   & 47/60 \\
Famous  & 9/30 \\
Mirror  & 28/30 \\
Unique  & 26/30 \\
Anonymized Basic   & 50/60 \\
Anonymized Famous  & 14/30  \\
Anonymized Mirror  & 26/30 \\
Anonymized Unique  & 27/40 \\
\bottomrule
\end{tabular}
\caption{Number of bugs noticed by the LLM grouped by test suite category.}
\label{tab:bugs-noticed-corrected}
\end{table}

\subsection{Is Implementation or Intent Specified?}
While Deepseek-R1 demonstrates the ability to detect bugs, an equally important point is how it reacts when a bug is found. In every reasoning trace we examined, the model included a statement indicating that its specification would be based on the provided implementation rather than any inferred intent. With example phrasing like

\begin{lstlisting}[style=cotstyle]
Wait, strlen returns the length, which is the number of characters excluding the null terminator. So for a string "aba", strlen is 3. So right starts at 3. But the valid indices are 0,1,2. So str[right] in the first iteration would be str[3], which is the null terminator. That's a problem. Oh, the original code is incorrect here. But the user's code is given, and we have to annotate it as is. So the function as written is wrong because it's comparing the null terminator. But the user's code may have a bug, but we have to add annotations based on the code as given.
\end{lstlisting}
\label{listing:palindrome1}

Based on this reasoning, Deepseek-R1 went on to create specifications that didn't match the intent of the user even though it was able to identify the intent of the code. One such specification can be seen in Fig.~\ref{code:paldindrome_impl}

\begin{figure}[hbt]
\begin{lstlisting}[style=cstyle]
/*@
  requires \exists integer i; i >= 0 && str[i] == '\0' && (\forall integer j; 0 <= j <= i ==> \valid_read(str + j));
  ensures \result == (\strlen(str) == 0);
  assigns \nothing;
*/
int isPalindrome(const char* str) {
    int left = 0, right = strlen(str); 
    while (left < right) {
        if (str[left] != str[right]) {
            return 0; // Not a palindrome
        }
        left++;
        right--;
    }
    return 1; // Palindrome
}
\end{lstlisting}
\caption{Specification generated for buggyu palindome.c when implementation is prioritized}
\label{code:paldindrome_impl}
\end{figure}

Although this default behavior may not align with the expectations of users hoping to generate specifications for intended behavior, it is easily addressed. By modifying the prompt to acknowledge the possibility of buggy code and to explicitly instruct the model to generate specifications for the intended behavior rather than the actual implementation, we were able to shift the model’s output accordingly. Specifications generated under these revised prompts more closely aligned with the intended functionality, even when the implementation contained subtle errors.

\begin{lstlisting}[style=cotstyle]
Wait, wait. Wait, in the code, right is initialized as strlen(str). But strlen returns the length of the string excluding the null terminator. So for a string like "abc", strlen is 3, indexes 0,1,2. So the initial right is 3, but then in the loop, str[right] would be accessing index 3, which is the null terminator. That's a bug. Because the first comparison would be between str[0] and str[3] (the null), which is wrong. So the code as written is incorrect. Because the right should start at strlen(str) - 1.

But according to the user's goals, if there's a bug, I should generate annotations for what the intended behavior is, not the implemented code. So the intended behavior is to check if the string is a palindrome correctly. So the function probably meant to set right to strlen(str) -1.

So the annotations should reflect the intended behavior, not the actual code. So I need to model what the correct function would require and ensure.
\end{lstlisting}
\label{reasoning:palindrome2}

Fig.~\ref{code:paldindrome_intent} shows the specification generated for our buggy palindrome program once the prompt is instructed to prioritize intent.  

\begin{figure}[hbpt]
\begin{lstlisting}[style=cstyle]
/*@
  requires \exists integer i; i >= 0 && str[i] == '\0' && (\forall integer j; 0 <= j <= i ==> \valid_read(str + j));
  ensures \result == (\strlen(str) == 0);
  assigns \nothing;
*/
int isPalindrome(const char* str) {
    int left = 0, right = strlen(str); 
    while (left < right) {
        if (str[left] != str[right]) {
            return 0; // Not a palindrome
        }
        left++;
        right--;
    }
    return 1; // Palindrome
}
\end{lstlisting}
\caption{Specification generated for buggyu palindome.c when intent is prioritized}
\label{code:paldindrome_intent}
\end{figure}

\subsection{Summary: Implementation vs. Intent}

Our qualitative analysis shows that Deepseek-R1, equipped with reasoning capabilities, is able to infer the intent behind programs even when presented with unfamiliar code, anonymized identifiers, and no documentation. It consistently demonstrated the ability to detect purposely subtle bugs-errors designed to evade casual inspection and reflected meaningfully on how to respond to such discrepancies. In every instance where a bug was identified, the model referred back to the prompt to determine whether to align its specification with the observed implementation or the inferred intent. While this LLM-based solution should be treated as the function \textbf{approximation} that is, LLMs such as Deepseek-R1 are a promising tool for specification synthesis, especially when guided by well-formed prompts and further model tuning.

\section{Augmenting LLM-based Specification Synthesis with Symbolic Analysis}
\label{experiment2}

Two common criticisms of LLM-based specification synthesis are: (1) the space of possible specifications is effectively unbounded, possibly leading to specifications that are trivially true (e.g., \texttt{requires true}) or simply incorrect; and (2) LLMs lack the precision provided by domain-specialized symbolic tools.

However, there is no inherent reason why LLM-based generation and symbolic analysis could not be combined.
%In fact, there are many promising avenues for combining these approaches.
In this work, we run symbolic tools on a given program and add their output to the LLM prompt. This method preserves the flexibility of the LLM while injecting the precision and focus of symbolic methods.

While many symbolic tools exist, we focus on two from the Frama-C verification framework: PathCrawler, which performs test generation through path exploration, and EVA, which performs static value analysis.

In this section, we aim to answer the following questions:
\begin{itemize}
  \item How do LLMs such as Deepseek-R1 interact with the output of symbolic tools?
  \item Do the specifications generated by LLMs change when symbolic output is included in the prompt?
  \item What are some plausible directions for future integrations of symbolic methods and LLM-based synthesis?
\end{itemize}

\subsection{Dataset}
For our study, we have chosen to utilise the 55 programs from the Pathcrawler test suite, which we will refer to as \textbf{pathcrawler\_tests}. Note that the test suite is not available online; it was provided to us by the Pathcrawler developers. Thereby, we can assume that this test suite was not directly used in the LLM's training (although it might have seen similar ones). This suite includes a variety of program types, balancing well-known algorithms like Binary Search with more niche programs such as a Soup Heater controller. It also contains small, specially crafted programs designed to test specific capabilities of Pathcrawler, adding another layer of diversity to our tests.

According to the Pathcrawler developers, the 55 programs are supposedly correct, in the sense that they are believed to correspond to their respective intention, and have no known bugs. Consequently, we examine with this suite to which extent our method produces accurate annotations for supposedly correct programs.

\subsection{Generating Annotations for Pathcrawler Tests}
\label{pc_generation_annotations}
In addition to all of the processes described in~\ref{section:generating_annotations}, we generate additional sets of annotations in for pathcrawler\_tests. For each program, we generate three sets of ACSL annotations: 
 \begin{enumerate}
     \item \textbf{baselines\_set}: Specifications generated using just the program in the prompt
      \item \textbf{pathcrawler\_set}: Specifications generated by including a compact representation of test-cases generated by Patchcrawler in the prompt.
     \item \textbf{eva\_set}: Specifications generated by running EVA on the program and including its report on potential value errors in the prompt.
 \end{enumerate}

\subsection{Evaluation}

\subsubsection{Annotation Counts}
First we quantify the annotations generated across all programs for our three sets: \textbf{baseline\_set}, \textbf{pathcrawler\_set}, and \textbf{eva\_set}. While this approach does not say anything about the quality or semantics of any annotations, it does provide us with a macro-level view of which kinds of annotations are being generated and at what frequency. This systematic approach allows us to zoom out and capture the influence of different symbolic contexts on the annotation generation process.

\subsubsection{Qualitative Analysis}
Similar to our analysis in Section~\ref{sec:intent_eval}, this section focuses on a qualitative evaluation of the specifications generated by the LLM, as well as the reasoning that precedes them. Our goal is not to assign a quantitative score to specification quality, but rather to understand the model's behavior, especially in response to symbolic analyses added to the prompt.

In this phase, we place particular emphasis on comparing specifications produced in the \texttt{baseline\_set} with those in the \texttt{pc\_set} and \texttt{eva\_set}, which include symbolic pre-analysis generated by tools such as PathCrawler and EVA. These comparisons allow us to evaluate how the presence of symbolic analysis affects the content of generated specifications.

Additionally, we analyze the model's reasoning steps prior to specification generation. This provides insight into how the symbolic analysis outputs are being interpreted and incorporated into the reasoning process that proceeds generation. This helps us to understand whether the analysis results are being used merely as contextual information or as direct cues.

\subsection{Counting Annotations}
In this section, we count the number of ACSL annotations produced in the specifications across the three datasets described in~\ref{pc_generation_annotations}. Specifically, we count occurrences of three core annotation types: \texttt{requires} clauses (preconditions), \texttt{ensures} clauses (postconditions), and \texttt{assigns} clauses (which describe memory mutations).

While a small number of other annotation types such as \texttt{logic} predicates or \texttt{ghost} variables were occasionally generated, they were too infrequent be worth including in our analysis. Our goal is to determine whether the inclusion of symbolic tools like PathCrawler leads to any noticeable differences in the number or type of annotations produced by the model.

\begin{figure*}[htbp]
\includegraphics[width=\textwidth]
{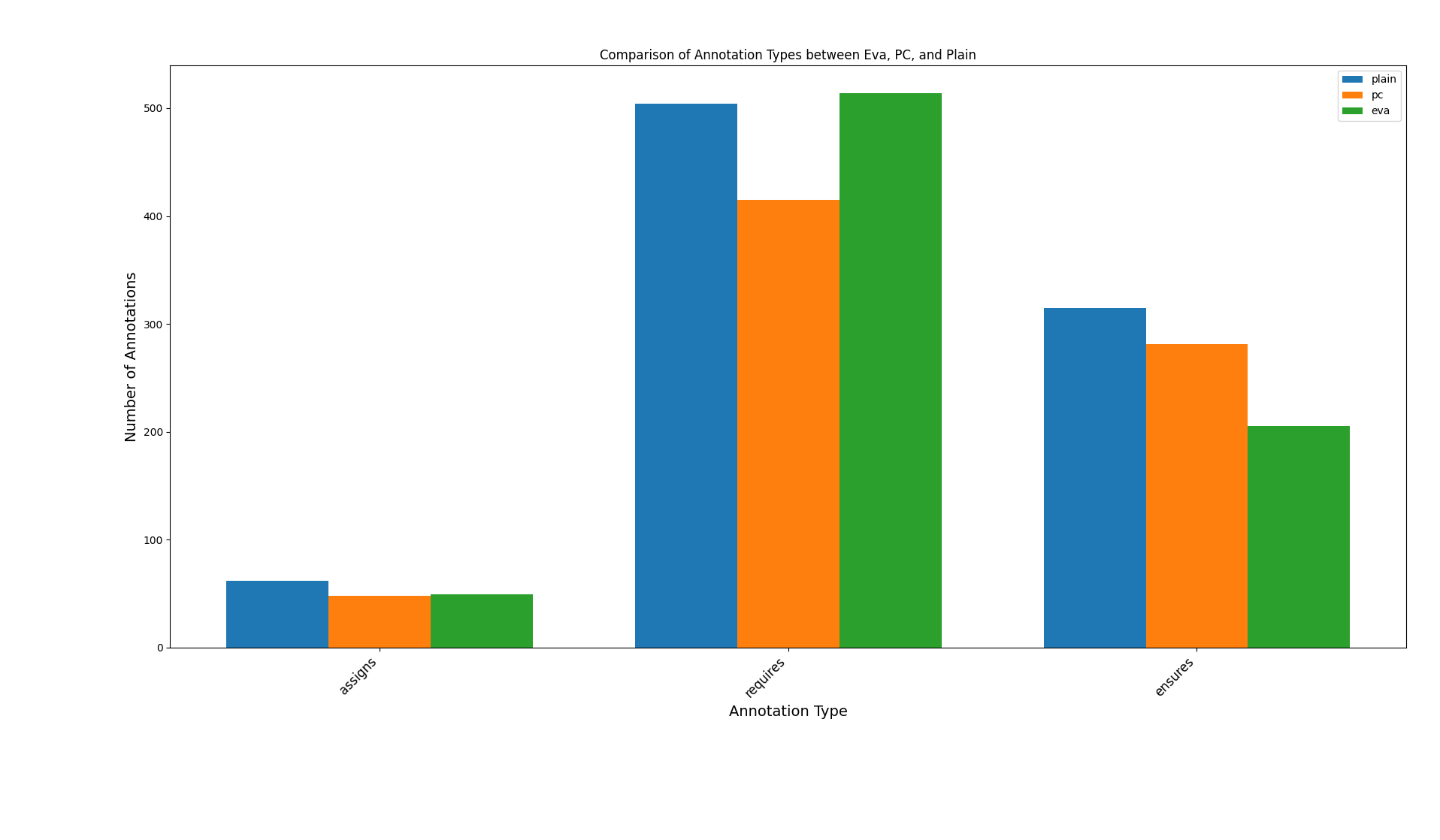}
\caption{Annotation-type counts for each prompt}
\Description{Bar chart comparing counts of preconditions, postconditions, and assigns clauses across different prompt configurations.}
\label{graphic:counts}
\end{figure*}

The counts are presented in Fig.~\ref{graphic:counts} on page~\pageref{graphic:counts}.

\subsection{Annotation Counts}

\subsubsection{More is Not Better}
An important point to consider when interpreting the annotation counts presented in Fig.~\ref{graphic:counts} on page~\ref{graphic:counts} is that a larger number of annotations does not necessarily indicate a better or more useful specification. For instance, a trivial postcondition such as \texttt{ensures true;}.

We encountered the issue of frivolous ACSL clauses in the previous iteration of this experiment, where GPT-4 often generated \texttt{assigns} clauses within the function body for every variable assignment. While these annotations adhered to ACSL syntax, they were semantically vacuous and did not aid in program verification or understanding.

When re-running our experiments with Deepseek, we observed that specifications generated without symbolic assistance tended to contain the highest number of annotations overall. However, unlike in our earlier studies, Deepseek did not produce clearly frivolous specifications. Instead, we observed a modest increase of approximately 20\% in annotation counts.

Although the reasoning traces did not provide a definitive explanation for this behavior, one plausible interpretation is that, in the absence of additional guidance from symbolic tools, the model focused more on easily identifiable elements of the code. This may have led to an overproduction of annotations reflecting basic implementation details.

\subsubsection{EVA and Preconditions}
The only annotation type for which we observed a clear increase was the number of preconditions generated when EVA reports were included in the prompt. This finding aligns with results previous iterations of this experiments where GPT-4 converted EVA alarms into preconditions to remedy possible runtime errors. We explore this behavior in more detail in the dedicated section on EVA that follows.

\subsection{PathCrawler}

\subsubsection{Usage of Examples}
When asked to generate specifications, Deepseek-R1 consistently reasons about code using examples whether or not they are explicitly provided in the prompt. As shown in Fig.~\ref{reasoning:adpcm_pc}, in the absence of examples the model invents its own to support its understanding of program behavior. However, when concrete examples are included in the prompt as is the case when using outputs from PathCrawler thostyle=csse provided examples are directly used during the reasoning process. This can be seen in the reasoning process shown in Fig.~\ref{reasoning:adpcm_no_pc}.

\begin{figure}[hbpt]
\begin{lstlisting}[style=cotstyle]
The CSV input has some test cases. For example, input_t[0] and input_t[1] are -37 and 0, and valeur 0. The output is 0. Wait, in the function 'testme', the output is 'retour', which depends on the return value of 'g'. Let's trace that.

@@In the first test case: n=2, valeur=0, t is [-37,0]. So 'g' checks if 0 < t[0] (-37)? No. Then check if 0 >= t[1] (0), yes. So r is set to -1. Then 'testme' gets i=-1, so retour is 0. Which matches the output 0.

Another test case: input_n=2, input_valeur=-91, t[0]=0, t[1]=62. So in 'g', valeur is -91. -91 < t[0] (0), so r is -2. Then 'testme' sets retour to 0. Output is 0. Correct.

Third test case: n=2, valeur=0, t is [0,12]. In 'g', 0 is not less than t[0], which is 0. Then check if 0 >= t[1] (12), no. Then loop over i=0 (since n=2, loop runs once). Check if 0 >= t[0] (0) and < t[1] (12). Yes. So r=0. So 'testme' returns 1. Which matches the output 1.@@

So from these test cases, the 'g' function seems to be looking for the index where 'valeur' is in the interval [t[i], t[i+1}), but only if the array is sorted in ascending order. Wait, in the third test case, t is [0,12], sorted. But what if the array isn't sorted? Let's say t is [5,3]. If valeur is 4, then t[0] is 5, which is larger than 4, so r stays -2. Then check if 4 >= t[1] (3), yes, so r is -1. The loop would check i=0, 4 >=5? No. So r remains -1. So 'g' returns -1, and 'testme' returns 0. So 'g' might expect that the array is sorted. But the code itself doesn't enforce that. So maybe the function 'g' is intended to work on a sorted array. So maybe the preconditions require that the array is sorted in ascending order.
\end{lstlisting}
\caption{Snippet of reasoning done for ADPCM.c (Adaptive differential pulse-code modulation) when provided with pathcrawler output}
\label{reasoning:adpcm_pc}
\end{figure}

\begin{figure}[hbpt]
\begin{lstlisting}[style=cotstyle]
But perhaps we can model the behavior more precisely. Let's think about the possible cases for 'testme':

- If g returns -1 or -2, retour is 0.
- If g returns between 0 and n-1, retour is 1.
- Else, retour is -1. But according to 'g's code, this case can't happen. So maybe there's an error here. But the code as written includes the else, so in ACSL, we need to cover all possibilities.
\end{lstlisting}
\caption{Snippet of reasoning done for ADPCM.c (Adaptive differential pulse-code modulation) when no input/output examples are provided}
\label{reasoning:adpcm_no_pc}
\end{figure}

This behavior is desirable. Unlike ad-hoc examples generated by the LLM, the input/output pairs produced by PathCrawler are selected based on structural analysis of the program, and the criteria for their generation can be configured and understood by the user interacting with PathCrawler. While the broader problem of structural test case generation remains an open research area, it is also mature enough to offer practical tools and workflows. Integrating such examples into the LLM’s input gives users a controllable way to guide the specification generation process with known and analyzable test scenarios.

\subsubsection{Non-trivial Example Reasoning}
The reasoning traces generated by Deepseek-R1 reveal that its engagement with input examples is non-trivial. Rather than merely accepting or describing examples, the model actively traces the inputs through the program logic and compares the outputs with its internal model of how the program works. This kind of step-by-step evaluation is distinctive to reasoning-style LLMs and was not observed in our previous experiments with GPT-4's chain-of-thought outputs.

Take, for example, Fig.~\ref{code:shift_pc} and its accompanying reasoning in Fig.~\ref{reasoning:shift_pc}. Deepseek does not stop at tracing examples; it also generalizes from them. In this case, the model traces multiple concrete input values through the bitwise logic of the program, including large negative integers. It performs bitwise shifts and masks, demonstrating an ability to simulate low-level operations "non-standard" numbers to verify behavior.

More notably, the model then abstracts these examples into general rules that relate specific input ranges to program outputs. This behavior is mirrored in both its reasoning trace and in the unprompted comments that it embedded in the specification. These annotations describe how different intervals of the input variable a influence the return value, reflecting a higher-level understanding of the program's semantics beyond what is visible in the code alone.

\begin{figure}[hbpt]
\begin{lstlisting}[style=cstyle]
/*@ 
  // Case when input is -1 gives exactly 5
  ensures a == -1 ==> \result == 5;
  // Values where right-shifted a is < -4 or >4 yield 2 (excluding a=-1)
  ensures (a != -1) && ((a >> 1) < -4 || (a >> 1) > 4) ==> \result == 2;
  // Right-shifted a in [-4,-2] or [1,4] gives 3 (excluding a=-1)
  ensures (a != -1) && ((a >> 1) >= -4 && (a >> 1) <= -2) ==> \result == 3;
  ensures (a != -1) && ((a >> 1) >= 1 && (a >> 1) <= 4) ==> \result == 3;
  // Right-shifted a in [-1,0] gives maximum non-special value 4
  ensures (a != -1) && ((a >> 1) >= -1 && (a >> 1) <= 0) ==> \result == 4;
*/
int testme(int a){
  int result = 0;
  if ((a >> 1) >= -4)
    result++;
  if ((a >> 1) <= 4)
    result++;
  if ((a >> 1) > -2)
    result++;
  if ((a >> 1) <= 0)
    result++;
  if (a == -1)
    result++;
  return result;
}
\end{lstlisting}
\caption{Specification generated for TestShiftRT}
\label{code:shift_pc}
\end{figure}

\begin{figure}[hbpt]
\begin{lstlisting}[style=cotstyle]
Let's analyze each case:
- For a=0:
  - (0 >> 1) is 0. So all four shift-related conditions: 0 >=-4 (yes), 0 <=4 (yes), 0 >-2 (yes), 0 <=0 (yes). So result is 4. Then the fifth condition a==-1 is false. So total 4, which matches the CSV.
- For a=-1:
  - (-1 >> 1) in C is implementation-defined, but typically arithmetic shift, so for signed integers, right shift of a negative number fills with 1s. So -1 in binary is all 1s. Shifting right by 1 gives -1 again. So (a >>1) is -1.
  - Check conditions:
    1. -1 >=-4 -> yes -> +1
    2. -1 <=4 -> yes -> +1
    3. -1 >-2 -> yes -> +1 (since -1 is greater than -2)
    4. -1 <=0 -> yes -> +1. So four from shifts. Then a==-1 -> +1. Total 5, which matches CSV.
... much more reasoning
1. (a>>1) >=-4 -> -830 million >=-4 -> yes. So +1.
2. (a>>1) <=4 -> -830 million <=4 -> yes. So +1. Now result is 2.
3. (a>>1) >-2 -> -830 million >-2 -> yes? Wait, no. -830 million is less than -2. So this condition is false. So no increment.
4. (a>>1) <=0 -> yes, since it's negative. So +1. Now result is 3?
... more reasoning
So from the code and data, we can see that:

- When (a>>1) is in [-4,4], but depending on the other conditions.

But the conditions are not all overlapping. For example:

Condition1: (a>>1) >=-4 -> possible values from -4 to MAX (for signed int)

Condition2: (a>>1) <=4 -> possible values from MIN to 4.

So the intersection of condition1 and condition2 is [-4,4].
\end{lstlisting}
\caption{Interesting snippets from the reasoning process being the specification generated for TestShiftRT with PathCrawler output in the prompt}
\label{reasoning:shift_pc}
\end{figure}

Comparing this to the specification generated without PathCrawler, shown in Fig.~\ref{code:shift_no_pc}, the specification generated with PathCrawler is both more desirable and semantically equivalent. While the PathCrawler-assisted version decomposes the specification into clear logical cases based on observed mappings of input values ranges to distinct outputs, the version generated without PathCrawler performs a more literal translation of the implementation into ACSL.

This direct translation is not technically incorrect, but it offers little interpretive value. In effect, it re-states the code in ACSL without offering additional insight. For a user trying to understand the behavior or verify the correctness of a program, they are essentially provided a guarantee that the program "does what it says"; they might as well read the code itself.

An observation, and maybe part of the true explanation for this behavior, across our examples is that reasoning without provided inputs - such as those provided by PathCrawler - leads the LLM to invest more effort into inventing examples on its own. This effort reduces the time and attention available for structural and logical generalization. In contrast, when supplied with example inputs, the model shifts more quickly to analyzing patterns and deriving more meaningful, generalized specifications.

\begin{figure}[hbpt]
\begin{lstlisting}[style=cstyle]
/*@ 
  @ ensures 0 <= \result <= 5;
  @ ensures \result == ((a >> 1 >= -4) + (a >> 1 <= 4) + (a >> 1 > -2) + (a >> 1 <= 0) + (a == -1));
*/
int testme(int a){
  int result = 0;
  if ((a >> 1) >= -4)
    result++;
  if ((a >> 1) <= 4)
    result++;
  if ((a >> 1) > -2)
    result++;
  if ((a >> 1) <= 0)
    result++;
  if (a == -1)
    result++;
  return result;
}
\end{lstlisting}
\caption{Specification generated for TestShiftRT with no PathCrawler output in the prompt}
\label{code:shift_no_pc}
\end{figure}

\subsubsection*{Input Pattern Recognition}

Another behavior we observed was Deepseek's ability to identify and leverage patterns in input data during its reasoning process.  The LLM appeared to actively search for consistent properties in the input such as ordering, bounds, or structure in order to infer preconditions or better understand the underlying logic of the program.

For example, in the PathCrawler output and accompanying reasoning showed in Fig.~\ref{csv:bsearch}, the model quickly picked up on the fact that all the provided inputs were sorted. This recognition was not only mentioned in its reasoning but also incorporated into the preconditions of the generated specification.

\begin{figure}[hbpt]
\begin{lstlisting}[style=csvstyle]
input_elem,input_A[0],input_A[1],input_A[2],input_A[3],input_A[4],input_A[5],input_A[6],input_A[7],input_A[8],input_A[9],output,verdict
94,0,0,0,0,0,0,0,0,0,0,0,success
7,0,0,0,0,0,0,0,0,0,7,1,success
0,0,0,0,0,0,0,0,0,0,0,1,success
\end{lstlisting}

\begin{lstlisting}[style=cotstyle]
For example, the third test case has all zeros, which is trivially sorted. The second test case's array has zeros except the last element 7, which is sorted. The first test case's array is all zeros. So, perhaps the function expects the array to be sorted. But the code doesn't enforce this. However, for the binary search to work correctly, the array must be sorted. Therefore, a requires clause in ACSL should state that the array is sorted
\end{lstlisting}
\caption{PathCrawler output for Bsearch.c}
\label{csv:bsearch}
\end{figure}

A more subtle instance of this Deepseek's ability to pick up on abstract patterns is illustrated in the PathCrawler output and accompanying reasoning showed in Fig.~\ref{csv:bugkpath}. There, Deepseek used a single example early in its reasoning trace to hypothesize a high-level intent for the program. This could factor into our observation that programs with pathcrawler output provided seemed to produce specifications with more abstract properties.

\begin{figure}[hbpt]
\begin{lstlisting}[style=csvstyle]
input_len,input_buffer[0],input_buffer[1],input_buffer[2],input_buffer[3],input_msg[0],input_msg[1],output,verdict
2,0,0,0,0,0,0,0,0,unknown
3,0,0,0,0,10,46,0,0,46,46,unknown
2,0,0,0,0,10,46,10,46,46,unknown
2,0,0,0,0,10,0,10,0,unknown
\end{lstlisting}

\begin{lstlisting}[style=cotstyle]
Next, looking at the CSV data. The columns are input_len, input_buffer (positions 0-3), input_msg (positions 0-1), output, verdict. The output seems to show the buffer after processing. For example, in the third row, input_msg is 10 and 46 (ASCII for '\n' and '.'), and the output becomes 10, 46, 46. This suggests that when "\n." is encountered, the code adds an extra '.' to the buffer, hence three characters: '\n', '.', '.'.
\end{lstlisting}
\caption{PathCrawler output and reasoning for BugKPath.c}
\label{csv:bugkpath}
\end{figure}

\subsubsection*{Small Numbers of Test Cases}
One immediately noticeable trend in our experiments was that programs with only a small number of test cases, typically those with a single reachable branch, gained little benefit from PathCrawler output. In these cases, the specification generated by Deepseek was essentially the same as that generated without symbolic assistance. This tracks as PathCrawler provided only minimal new information to the model.

Even when there were multiple paths for PathCrawler to generate inputs for, having only one test case per control path occasionally led Deepseek to incorrect conclusions. The lack of variation in the data left the model with insufficient information to accurately generalize behavior. Going back to our previous example, in the specification generated for the program \texttt{BugKPath}, shown in Fig.~\ref{code:bugkpath_pc} along with its reasoning in Fig.~\ref{reasoning:bugkpath_pc}, Deepseek inferred that a transformation only occurred at the beginning of a string. In reality, \texttt{BugKPath} was intended to replace every occurrence of "\textbackslash n." in a buffer with "\textbackslash n..". The model’s misunderstanding stemmed from the fact that the only test case it was given featured the target pattern at the start of the input. Without additional examples, the model generalized prematurely, producing a specification that reflected the limited input rather than the actual behavior.

\begin{figure}[hbpt]
\begin{lstlisting}[style=cotstyle]
Additionally, from the CSV data, when the input message starts with "\n.", the buffer contains "\n..". An ensures clause could capture this, but in ACSL, it's challenging to relate the input msg to the output buffer for all cases. However, based on the example, we can add an ensures clause that if the first two characters are "\n.", then the buffer's first three characters are "\n..". But this is very specific. For example:

ensures (len >= 2 && msg[0] == '\n' && msg[1] == '.') ==> 
         (buffer[0] == '\n' && buffer[1] == '.' && buffer[2] == '.');

But this is just one case. It's unclear if this is desired. However, the user's examples show that the code adds an extra '.', so the ensures clause could mention this.
\end{lstlisting}
\caption{PathCrawler output and reasoning for BugKPath.c}
\label{reasoning:bugkpath_pc}
\end{figure}

\begin{figure}[hbpt]
\begin{lstlisting}[style=cstyle]
/*@ 
  requires len >= 0;
  requires (len > 0) ==> \valid_read(msg + (0..len-1));
  requires \valid(buffer + (0..BUFSZ-1));
  assigns buffer[0..BUFSZ-1];
  ensures (len >= 2 && \old(msg[0]) == '\n' && \old(msg[1]) == '.') ==> 
          (buffer[0] == '\n' && buffer[1] == '.' && buffer[2] == '.');
*/
void testme (char *msg, int len, char *buffer)
{
  int i;
  int j;
  //char buffer[BUFSZ];
  int limit = BUFSZ - 1;

  for (i = 0; i < len; ) {
    for (j = 0; i < len && j < limit; ){
      
      if (i + 1 < len 
          && msg[i] == '\n' 
          && msg[i+1]== '.') {

        buffer[j] = msg[i]; /* Suppose j == limit - 1 */
        j++;
        i++;
        
        buffer[j] = msg[i]; /* Now j == limit */
        j++;
        i++;
        /* BAD */

        buffer[j] = '.';    /* Now j == limit + 1 = sizeof(buffer) */
        j++;
      } else {
        buffer[j] = msg[i];
        j++;
        i++;
      }
    }
  }
}
\end{lstlisting}
\caption{PathCrawler output and reasoning for BugKPath.c}
\label{code:bugkpath_pc}
\end{figure}

\subsubsection{PathCrawler Summary}
Including PathCrawler output into the specification synthesis process had a subtle but noticeable effect on the quality of the generated specifications. In particular, specifications generated with PathCrawler displayed more abstract reasoning and better structural organization when "high quality" input/output examples were provided.

This improvement appeared to result from Deepseek's reasoning process being freed from the need to generate examples itself, allowing it to concentrate more on generalizing program behavior. High quality examples were defined as those that covered all execution paths with multiple examples per path. Without full path coverage, the control flow of the program cannot be accurately captured. Without sufficient examples per path, the model risks producing specifications tailored to a single example.

PathCrawler, as a tool, was designed to produce the minimal number of test cases required to achieve full path coverage. While this goal aligns well with test generation, it is not necessarily optimal for specification synthesis. However, there is no reason why PathCrawler or a similar tool could not be extended or adapted to better suit the needs of specification synthesis. More broadly, providing LLMs with curated input/output examples formatted for both human readability and alignment with the underlying code appears to be a promising direction for guiding specification synthesis in practice.

\subsection{Prompt with EVA annotations}
EVA, a static analysis tool, specialises in conducting a comprehensive value analysis of C programs. The report that it generates includes the detailed outcomes of the value analysis but also lists alarms that signify possible runtime errors linked to these value states. Such alarms are indicators of conditions under which the program might fail or behave unexpectedly, essentially flagging areas of the code that are prone to errors due to specific input values or execution paths.

\subsubsection{Avoiding Runtime Errors}
In our analysis of the \textbf{eva\_set}, a distinct characteristic emerges: the prevalence of precondition annotations that serve to help the function avoid runtime errors. These preconditions appear to be a direct result of the runtime alarms included the EVA reports. The alarms detail problematic input ranges, such as values that could cause the program to behave unpredictably or fail -- common issues highlighted include index-out-of-bound errors, divide-by-zero errors, and integer overflows.

An example of these value domains can be seen in the specification generated for the program \textbf{Alias5}, seen in Fig.~\ref{code:alias5_eva}. The specification clearly aligns with the EVA report included in its synthesis prompt.

\begin{lstlisting}[style=cotstyle]
...
[eva:alarm] temp_files/tmphpdyn83w/eva_temp.c:8:Warning:
  signed overflow. assert -2147483648 <= x * 2;
[eva:alarm] temp_files/tmphpdyn83w/eva_temp.c:8:Warning:
  signed overflow. assert x * 2 <= 2147483647;
[eva:alarm] temp_files/tmphpdyn83w/eva_temp.c:9:Warning:
  signed overflow. assert -2147483648 <= v - y;
[eva:alarm] temp_files/tmphpdyn83w/eva_temp.c:9:Warning:
  signed overflow. assert v - y <= 2147483647;
[eva:alarm] temp_files/tmphpdyn83w/eva_temp.c:11:Warning:
  out of bounds write. assert \valid(tab + 2);
[kernel] temp_files/tmphpdyn83w/eva_temp.c:11:Warning:
  all target addresses were invalid. This path is assumed to be dead.
\end{lstlisting}

\begin{figure}[htbp]
\begin{lstlisting}[style=cstyle]
/*@ 
  requires -1073741824 <= x <= 1073741823; // Prevent x*2 overflow
  requires -2147483648 <= (x*2 - y) <= 2147483647; // Prevent v-y overflow
  requires \valid(tab + (0..2));            // Ensure tab[2] is accessible
  requires \valid(tab + (y +4));            // Ensure tab[y+4] access is valid
  requires \valid(tab + (2 + (x*2 - y)));   // Ensure *(pt+v) write is valid
*/
int testme(int x, int y, int tab[]) {
  int v;
  int *pt;
  int **ppt;
  v = x * 2;
  v = v - y;        /* v = (x * 2) - y */
  pt = &tab[2];
  tab[2] = x;
  ppt = &pt;
  pt = pt + 1;      /* pt = &tab[2] + 1 */
  *ppt = &tab[2];   /* pt = &tab[2] */
  *(pt + v) = y;    /* tab[2 + (x * 2) - y] = y */
  if (tab[y + 4] > 5)
    return 1;
  return 0;
}
\end{lstlisting}
\caption{Alias5 specification generated with an EVA report as context}
\label{code:alias5_eva}
\end{figure}

Notice how the alarms are utilised by the LLM in generating preconditions that ensure that the program avoids both integer overflow and index-out-of-bounds errors. These precise runtime-aware annotations are largely unique to annotations generated with EVA as the reports provide a level of direction and precision absent in the baseline prompt.

\subsubsection{EVA Tunnel-vision}
In examining the impact of incorporating EVA reports into the specification generation process, it becomes apparent that the LLM prioritises the analysis of the EVA report over the direct reasoning about the program's implementation. This observation emerges distinctly when analysing the specifications and reasoning CoT produced by the LLM. For example, compare the specifications seen in Fig.~\ref{code:tritype_eva} and Fig.~\ref{code:tritype_baseline} where the former represents a specification generated with an EVA report in the prompt and the latter without. The specification generated in Fig.~\ref{code:tritype_eva} completely ignores all of the structure and logic that was focused on normally and instead completely focuses on eliminating runtime errors. This is also reflected in the reasoning about the EVA report displayed in Fig.~\ref{listing:cot2} on page~\pageref{listing:cot2}.

\begin{figure}[hbtp]
\begin{lstlisting}[style=cstyle]
/*@ 
  requires INT_MIN <= i + j <= INT_MAX;
  requires INT_MIN <= i + k <= INT_MAX;
  requires INT_MIN <= j + k <= INT_MAX;
  ensures \result == 1 || \result == 2 || \result == 3 || \result == 4;
*/
int testme(int i, int j, int k){
  int type_code;
  if ((i == 0) || (j == 0) || (k == 0)) type_code = 4;     /* line 13 */
  else {
    type_code = 0;
    if (i == j) type_code = type_code + 1;                 /* line 16 */
    if (i == k) type_code = type_code + 2;                 /* line 17 */
    if (j == k) type_code = type_code + 3;                 /* line 18 */
    if (type_code == 0){                                   /* line 19 */
      if ((i+j <= k) || (j+k <= i) || (i+k <= j))          /* line 20 */
	type_code = 4;
      else
	type_code = 1;
      }
    else if (type_code > 3) type_code = 3;                 /* line 25 */
    else if ((type_code == 1) && (i+j > k)) type_code = 2; /* line 26 */
    else if ((type_code == 2) && (i+k > j)) type_code = 2; /* line 27 */
    else if ((type_code == 3) && (j+k > i)) type_code = 2; /* line 28 */
    else type_code = 4;
    }
  return type_code;
}
\end{lstlisting}
\caption{Tritype specification generated with an EVA report as context}
\label{code:tritype_eva}
\end{figure}

\begin{figure}[hbpt]
\begin{lstlisting}[style=cstyle]
/*@ 
  ensures 1 <= \result <= 4;
  ensures \result == 4 <==> (i == 0 || j == 0 || k == 0 || (i + j <= k) || (j + k <= i) || (i + k <= j));
  ensures \result == 3 <==> (i != 0 && j != 0 && k != 0 && i == j && j == k);
  ensures \result == 2 <==> (i != 0 && j != 0 && k != 0 && 
    ((i == j && i + j > k) || (i == k && i + k > j) || (j == k && j + k > i)));
  ensures \result == 1 <==> (i != 0 && j != 0 && k != 0 && 
    i != j && i != k && j != k && (i + j > k) && (j + k > i) && (i + k > j));
*/
int testme(int i, int j, int k){
  int type_code;
  if ((i == 0) || (j == 0) || (k == 0)) type_code = 4;     /* line 13 */
  else {
    type_code = 0;
    if (i == j) type_code = type_code + 1;                 /* line 16 */
    if (i == k) type_code = type_code + 2;                 /* line 17 */
    if (j == k) type_code = type_code + 3;                 /* line 18 */
    if (type_code == 0){                                   /* line 19 */
      if ((i+j <= k) || (j+k <= i) || (i+k <= j))          /* line 20 */
	type_code = 4;
      else
	type_code = 1;
      }
    else if (type_code > 3) type_code = 3;                 /* line 25 */
    else if ((type_code == 1) && (i+j > k)) type_code = 2; /* line 26 */
    else if ((type_code == 2) && (i+k > j)) type_code = 2; /* line 27 */
    else if ((type_code == 3) && (j+k > i)) type_code = 2; /* line 28 */
    else type_code = 4;
    }
  return type_code;
}
\end{lstlisting}
\caption{Tritype specification generated with the baseline prompt}
\label{code:tritype_baseline}
\end{figure}

\begin{figure}[hbpt]
\begin{lstlisting}[style=cotstyle]
The provided C program is a function `testme` that takes three integer inputs `Side1`, `Side2`, and `Side3` and uses these inputs to calculate and return a value `triOut`.

From the EVA report, we can see several warnings about potential signed overflow in the program. This occurs when the sum of two sides of the triangle is close to the maximum value an integer can hold (`2147483647`), leading to a possible overfl
\end{lstlisting}
\caption{Reasoning performed about an EVA report run on Tritype.c}
\label{listing:cot2}
\end{figure}

In this case we notice that while the specification in \textbf{eva\_set} was able to infer that this program was a triangle classifier, its focus was clearly taken by the provided EVA report.

This observation raises concerns about LLM-based specification synthesizers' ability to balance the input from static analysis tools like EVA that provide clear direction for the LLM. While the safety and domain boundaries are well-captured, the essence of what the program is designed to do can sometimes be overshadowed by the focus on avoiding runtime errors and handling edge cases as dictated by the EVA report.

\subsubsection{EVA Summary}
The effects of adding EVA reports to our prompts were more clear than those of PathCrawler. EVA's analysis produced a direct mapping from its runtime alarms to ACSL preconditions, which the LLM translated consistently and reliably.

This raises a question: do we even need an LLM for this task? Since EVA's alarms can be mechanically interpreted as input restrictions, one could imagine writing a dedicated tool to convert them directly into \texttt{requires} clauses. To some extent, this is a valid point when the symbolic output aligns cleanly with the target specification language.

However, this assumes the existence of a purpose-built translation tool for every symbolic analyzer and every specification formalism used in practice. Such tools are rarely developed comprehensively and are unlikely to exist across the full combinatorial space of symbolic tools and specification languages.

Instead, what we demonstrate with EVA is how easily LLMs can flexibly incorporate the output of symbolic analysis without needing custom integration. In our case, we simply appended the raw EVA report to the LLM prompt, and the model responded with annotations that captured the relevant constraints effectively. This showcases the adaptability of LLM-based specification synthesis in heterogeneous toolchains.

\section{Related Work}

%\begin{itemize}
%    \item Wolfgang: code generation ("cannot cover here in general, but MIT work ...")
%    \item Contrast to work on code generation
%    \item May be: work on generating models for model driven SW development, like M.Wirsing, and B.Steffen.
%\end{itemize}

There is an ever-growing body of work exploring the opportunities of combining LLMs with various tools for formal-methods and theorem proving. 
In the domain of proof-assistants in particular, a few works have explored the task of synthesising properties or lemmas using neural methods \cite{urban2020,rabe2021,johansson2023exploring} with varying success. More focus has been on creating models for generating proofs, with applications to most mainstream proof assistants like Isabelle/HOL, Coq and Lean \cite{lemmanaid,yang2023leandojo,thor,baldur,tactician}.
%% "reasoning" models such as
Recent developments of proof co-pilots for Lean are ongoing, aiming to make next-tactic suggestions for the user while creating proof scripts \cite{welleck2023llmstep,song2024towards}. Meanwhile in Isabelle/HOL, research has been made on template-based lemma conjecturing \cite{lemmanaid}.

For contract-based verification tools, the majority of research on annotation generation has been centered on generating loop invariants (e.g., \cite{kamath2023finding,wehrheimAndCoFASE24, wu2024llm, specgen}) needed to verify correctness of programs wrt.\ a \emph{given} functional specification. There is also work on generating assertions by LLMs with previously encountered errors added to the input \cite{mugnier2024laurel}, to support search for correctness proofs. With the latter, we have in common that some result from symbolic analysis is used when, iteratively, prompting LLMs. However, these works assume a functional specification to be given. Our research, on the other hand, addresses largely the problem of \emph{inferring} a functional specification for a program. Another approach within specification synthesis has focused on using static analysis to decompose larger programs into smaller, more manageable units, allowing the LLM to reason about each segment individually without losing focus due to token limitations~\cite{enchanting}.

This ability to generate specifications based on inferred intent could be particularly useful for developer productivity and user adoption. Silva et.al.~\cite{silva2024leveraging} describe how LLMs can be leveraged to help less experienced users craft formal specifications and prove their correctness in the Dafny programming language. In contrast to the theorem proving experiments, the work with Dafny did not utilize a model fine tuned on any sort of Dafny or contract-based verification. Some finetuning of contract-based specifications in the Java ecosystem was performed by~\cite{greiner2024automated} where they were able to finetune a model based on a dataset compromised of thousands of java programs and their accompanying JML contracts.

\section{Limitations}
\subsection{LLMs}
LLMs remain a relatively new  technology, and obtaining reproducible results from them is still a somewhat open topic, particularly when accessed through hosted services such as Deepseek or OpenAI. These services often operate behind proprietary APIs, and it is difficult to know what internal updates or changes may be introduced over time. This introduces a potential source of variability in results that cannot be easily controlled or accounted for.

In this work, we employed elevated temperature settings to encourage more diverse and creative outputs from the model. While this approach improves specification coverage and variety, it also increases randomness and reduces reproducibility. There is a tradeoff here: the traits that make LLMs effective for synthesis, creativity and flexibility, also make them more difficult to evaluate.

Despite these limitations, we were generally able to reproduce, using Deepseek-R1, the overall findings of our earlier experiments conducted with GPT-4 \cite{granberry2024specify} (for as far as the experiments of this work overlap with the experiments in \cite{granberry2024specify}). However, it is important to note that our qualitative analysis could be subject to confirmation bias. Since we had prior expectations from earlier results, there is an inherent risk that we interpreted or evaluated outcomes through the lens of those expectations during this second round of analysis.

\subsection{Datasets}
While the Intent Dataset used in this work was fully handcrafted and tailored to our experimental objectives, the PathCrawler dataset was not. The suite of programs included in the PathCrawler Set was originally developed to exercise the capabilities and edge cases of the PathCrawler tool itself. As such, a significant portion of these programs were not designed with specification synthesis in mind and do not always lend themselves to interesting specifications.

The advantage of using the PathCrawler Set was that it came with infrastructure for symbolic execution already in place. Specifically, the programs included input constraint files and configuration settings that enabled PathCrawler to explore paths effectively and generate inputs within meaningful bounds. This allowed us to test the integration of symbolic tools without first needing to develop a large number of compatible test cases from scratch. Moreover, the PathCrawler Set was closed source, which limits the danger of LLMs being contaminated with this data prior to our experiments.

Nonetheless, it is important to recognize that the total dataset used in this study consisted of approximately 50 programs, and only a subset of these were structured in a way that allowed for particularly expressive specifications. This limits the generality of our findings and highlights the need for future work on curated datasets designed specifically for evaluating specification synthesis across a broader spectrum of program behavior.

\section{Future Work}

While there are many ways to expand on this work, we highlight three especially interesting directions: models, datasets, and tooling.

\textbf{Models.} We have focused on general-purpose large language models (LLMs) hosted by providers such as Deepseek (here) and OpenAI (in \cite{granberry2024specify}). Inspired by recent efforts in the theorem proving community to fine-tune models for lemma generation or tactic prediction, we aim to experiment with fine-tuning models specifically for contract-based deductive verification. This could involve training on large dataset of programs annotated with ACSL specifications or similar formal annotations, with the goal of improving the model’s ability to generate and reason about ACSL specifications. If this is successful, the method is likely to carry over to other specification languages,
e.g., JML for Java. 

\textbf{Datasets.} In early experiments, we attempted to incorporate verification as part of our evaluation. However, earlier LLMs struggled with syntactic correctness and often with generating necessary loop invariants. These limitations led to largely uninteresting results. Moving forward, we hope to design a dataset that includes a larger set of programs that are realistically verifiable, possibly focusing on real-world, loop-free examples to start with. Such a dataset would allow us to revisit the question of whether specification synthesis from LLMs can be effectively integrated into verification pipelines.

\textbf{Tooling.} Our current work integrates symbolic outputs using two tools from the Frama-C ecosystem: PathCrawler and EVA. However, many other tools exist in the formal methods community that could provide useful results for specification synthesis. Whether through prompt-level integration, preprocessing, or post-processing, we see potential in exploring how the output of these tools can be used to enhance LLM-based synthesis.

\section{Conclusion}

This paper addresses two core challenges in LLM-based specification synthesis: the high variability in generated specifications and the possibility of conflicting implementation and intent in a given program. Using Deepseek-R1, a reasoning-focused LLM, we found that the model is well-suited to reasoning about subtle discrepancies between what code does and what it is intended to do. Across a range of programs, Deepseek consistently identified syntactically subtle but semantically significant bugs and responded to prompt directives that emphasized the program’s intent over its buggy implementation.

To reduce the variability in specifications, we explored whether symbolic output from formal tools could guide the model toward generating more consistent and directed specifications. We appended symbolic information from two Frama-C tools, PathCrawler and EVA, directly to the prompts and observed changes in the resulting specifications. PathCrawler, by providing input/output examples, appeared to shift the model's attention toward structural correctness and abstract logical relationships. Deepseek often used these examples as a substitute for self-generated test cases, allowing it to focus more directly on program logic. However, the utility of this approach depended heavily on the quality and quantity of the generated test cases.

In contrast, EVA's static value analysis yielded a different effect. By identifying potential runtime errors through value analysis, EVA prompted Deepseek to generate preconditions that would prevent those errors, resulting in more safety-focused specifications. While this led to stronger guarantees about program safety, it often came at the expense of postconditions.

Together, these results suggest that neurosymbolic prompting provides a viable path toward controlling both the structure and focus of LLM-generated specifications. By influencing the synthesis process with formal tool outputs, we can steer generation toward specific kinds of properties, helping address key issues of alignment and variability.

This work is aligned to the vision of trustworthy triple copiloting of implementations, tests, and specifications (`TriCo'), as co-outlined by two of the authors in \cite{ahrendt2022trico}. However, in that vision paper, we focused on the bilateral relations of the three artefacts, whereas here, when adding the Pathcrawler output, we use implementations and tests at once when generating specifications. More generally, we see our work as a contribution to the more general aim of combining the complementary strengths of machine learning and exact analyses for effective and reliable development of trustworthy software.

\bibliographystyle{splncs04}
\bibliography{references}
\end{document}